\documentclass[10pt]{article}
\usepackage{fullpage}
\usepackage{amsmath}
\usepackage{amssymb}
\usepackage{amsfonts}
\usepackage[dvips]{epsfig}
\usepackage{color}
\usepackage{cite}

\def\##1{\underline{#1}}
\def\=#1{\underline{\underline{#1}}}

\def\+
#1{\underline{\bf #1}}
\def\*#1{\underline{\underline{\bf #1}}}

\def\r#1{(\ref{#1})}
\def\l#1{\label{#1}}
\def\c#1{\cite{#1}}

\def\le{\left(}
\def\ri{\right)}
\def\les{\left[}
\def\ris{\right]}
\def\lec{\left\{}
\def\ric{\right\}}

\def\.{\mbox{ \tiny{$^\bullet$} }}

\def\eps{\varepsilon}

\def\epso{\eps_{\scriptscriptstyle 0}}

\def\muo{\mu_{\scriptscriptstyle 0}}

\def\ko{k_{\scriptscriptstyle 0}}

\def\ux{\hat{\#u}_x}
\def\uy{\hat{\#u}_y}
\def\uz{\hat{\#u}_z}

\def\calA{{\cal A}}
\def\calB{{\cal B}}

\begin{document}

\begin{center}

\LARGE{ {\bf On electromagnetic surface waves supported by an isotropic chiral material
}}
\end{center}
\begin{center}
\vspace{10mm} \large

{\bf James Noonan}$^1$ and {\bf Tom G. Mackay}$^{1,2,}$\footnote{Corresponding author. E--mail: T.Mackay@ed.ac.uk}

\vspace{2mm}

$^1$\emph{School of Mathematics  and
   Maxwell Institute for Mathematical Sciences, \\University of Edinburgh, Edinburgh
EH9 3JZ, United Kingdom}

 \vspace{2mm}

 $^2$\emph{NanoMM---Nanoengineered Metamaterials Group, Department of Engineering Science and
Mechanics, Pennsylvania State University, University Park, PA
16802--6812, USA}

\vspace{5mm}
\normalsize

\end{center}

\begin{center}
\vspace{15mm} {\bf Abstract}
\end{center}

 Electromagnetic surface waves supported by an isotropic chiral material were investigated via the associated canonical boundary-value problem. 
 Specifically, two scenarios were considered: surface waves guided by the planar interface of an isotropic chiral material and (a) an isotropic dielectric  material 
and (b) a uniaxial dielectric (or plasmonic) material. Both plasmonic and non-plasmonic achiral partnering materials were investigated.
 In scenario (a) surface waves akin to surface-plasmon-polariton (SPP) waves were excited, while in scenario (b) 
surface waves akin to Dyakonov waves and akin to SPP waves were excited. For numerical studies, an isotropic chiral material capable of  simultaneously supporting attenuation and amplification of plane waves, depending upon circular polarization state, was used. The amplitude of the SPP-like waves
could be either amplified or attenuated, depending upon the relative permittivity of the isotropic dielectric partnering material
 for scenario (a), or depending upon the direction of propagation relative to the optic axis of the uniaxial dielectric partnering material for scenario (b).

\vspace{5mm}

\vspace{5mm}

\section{Introduction}

The planar interface of two dissimilar materials can guide the propagation of electromagnetic surface waves. A variety of different types of electromagnetic surface wave have been identified, with the type depending upon whether the partnering materials are isotropic or anisotropic, dissipative or nondissipative, homogeneous or nonhomogeneous,  and so forth \c{ESW_book}. The surface-plasmon-polariton (SPP) wave \c{Pitarke,Maier}~---~which is guided by the planar interface of a metal 
and a dielectric material~---~is the most  familiar type of electromagnetic surface wave, being widely exploited in optical sensing applications \c{Homola_book,AZLe}. The Dyakonov surface wave \c{Dyakonov88}~---~which is guided by the planar interface of an isotropic dielectric material and an anisotropic dielectric material~---~has also been widely reported upon \c{Walker98,Takayama_exp}. Dyakonov waves are promising for applications in optical communications \c{DSWreview}.

As compared to achiral materials, chiral materials \c{Beltrami}~---~with their inherent magnetoelectric coupling~---~offer wider opportunities for surface-wave propagation. However, 
whereas surface waves guided by  planar interfaces involving achiral partnering materials have been comprehensively studied, there have been only a few studies of the surface waves supported by  chiral materials \c{Pattanayak,Engheta,Fantino,Pellegrini}. And these few studies have largely focussed on isotropic partnering materials and  nondissipative chiral materials. In the present paper we address this issue by
investigating surface waves supported by a  chiral material with the effects of dissipation and anisotropy of the partnering 
materials being taken into account. In particular, a novel type of chiral material that can simultaneously support attenuation and amplification of plane waves \c{ML_JO}, depending upon circular polarization state, is considered. Our theoretical and numerical studies
are based on the canonical boundary-value problem for surface-wave propagation \c{ESW_book}.

In the following, the permittivity and permeability of free space are written as $\epso$ and $\muo$, respectively. The free-space wavenumber is $\ko = \omega \sqrt{\epso \muo}$, where $\omega$ is the angular frequency. The operators $\mbox{Re} \lec \. \ric$ 
and $\mbox{Im} \lec \. \ric$ deliver the
real and imaginary parts of  complex-valued quantities, and $i = \sqrt{-1}$. 
Single underlining signifies a 3 vector while double underlining  signifies a 3$\times$3 dyadic. 
The triad of unit vectors aligned with the Cartesian axes are denoted as $\lec \ux, \uy, \uz \ric$.


\section{Isotropic chiral  material/isotropic dielectric  material interface} \l{iso_sec}

Let us consider the canonical boundary-value problem for surface waves guided by the planar  interface of an isotropic chiral  material
and an isotropic dielectric  material. Both partnering materials are homogeneous.
The isotropic chiral  material, labeled $\mathcal{A}$, fills
 the half-space $z>0$ and is characterized by the frequency-domain Tellegen constitutive relations \c{Beltrami}
\begin{equation} \l{cr}
\left.
\begin{array}{l}
 \#D (\#r) = \epso \eps_\mathcal{A} \#E (\#r) + i \sqrt{\epso \muo}  \xi_\mathcal{A} \#H (\#r) \,\\ [5pt]
 \#B (\#r)= - i \sqrt{\epso \muo} \xi_\mathcal{A} \#E (\#r) + \muo \mu_\mathcal{A} \#H (\#r) \,
\end{array}
\right\}\, \qquad z>0 .
\end{equation}
The relative permittivity scalar $\eps_\calA$, the relative  permeability scalar $\mu_\calA$, and the relative
chirality pseudoscalar
$\xi_\calA$ are frequency dependent and complex valued, per the principle
of causality embodied by the Kramers--Kronig relations \cite{2lines}.
The isotropic dielectric  material, labeled $\mathcal{B}$, fills
 the half-space $z<0$ and is characterized by the relative permittivity 
$\eps_\mathcal{B}$.

\subsection{Theory}

The electromagnetic field phasors in the  partnering materials $\calA$ and $\calB$ are represented by
\begin{equation} \label{planewave}
\left.\begin{array}{l}
 \#E_{\,\ell} (\#r)=  \#{\mathcal E}_{\,\ell} \,\exp\left({i\#k_{\,\ell} \cdot\#r}\right) \\[4pt]
\#H_{\,\ell}(\#r)= \#{\mathcal H}_{\,\ell} \,\exp\left({i\#k_{\,\ell} \cdot\#r}\right)
 \end{array}\right\}\,, \qquad   \ell \in\left\{ \calA, \calB \right\}\,.
\end{equation}
The  amplitude vectors $ \#{\mathcal E}_{\,\ell} $ and $ \#{\mathcal H}_{\,\ell}$ have complex-valued components, and so does
the wave vector
 $\#k_{\,\ell}$.  The field phasors (and the wave vector)   can vary with angular frequency $\omega$. Without loss of generality, we consider the  surface-wave  propagation parallel to $\ux$ in the $xy$ plane; i.e., $\uy\cdot\#k_{\,\ell}\equiv 0$.


In the  half-space $z>0$, the Maxwell curl postulates  yield
\begin{equation} \l{MP_A}
\left.
\begin{array}{l}
\#k_{\, \calA} \times \#{\mathcal E}_{\,\calA} - \omega  \le - i \sqrt{\epso \muo} \xi_{\calA}  \#{\mathcal E}_{\,\calA} +
\muo \mu_\mathcal{A} \#{\mathcal H}_{\,\calA} \ri = \#0
\vspace{4pt} \\
\#k_{\, \calA} \times \#{\mathcal H}_{\,\calA} + \omega \le  \epso \eps_{\calA}  \#{\mathcal E}_{\,\calA}
+ i \sqrt{\epso \muo} \xi_{\calA}  \#{\mathcal H}_{\,\calA}
\ri = \#0
\end{array}
\right\}\,,
\end{equation}
where the wave vector
\begin{equation} \l{wvA}
\#k_{\, \calA} = \ko \le q \, \ux + i \alpha_\calA \uz \, \ri
\end{equation}
and ${\rm Re}\lec\alpha_\calA\ric>0$   for surface-wave propagation.
On combining  Eqs.~\r{MP_A} and Eq.~\r{wvA}, a biquadratic dispersion relation
emerges for $\alpha_\calA$. The two   $\alpha_\calA$  roots with non-negative real parts are identified as
\begin{equation} \l{a_decay_const}
\left.
\begin{array}{l}
\alpha_{\calA 1} =  \sqrt{q^2 - \kappa_R^2  } \vspace{8pt}\\
\alpha_{\calA 2} =
 \sqrt{q^2 - \kappa_L^2  }
\end{array}
\right\}\,,
\end{equation}
with the complex-valued scalars
\begin{equation}
\left.
\begin{array}{l}
\kappa_R=  \sqrt{\eps_\calA \mu_\calA } + \xi_\calA  \vspace{6pt}\\
\kappa_L=  \sqrt{\eps_\calA \mu_\calA } - \xi_\calA  \end{array} \right\}
\end{equation}
being associated with the relative wave numbers for right and left circularly-polarized light in an unbounded chiral medium \c{Beltrami}.
Accordingly the field-phasor amplitudes are given as
\begin{equation}
\left.
\begin{array}{l} \l{A_p}
\#{\mathcal E}_{\,\calA} = A_{\calA 1}\, \#{\mathcal E}_{\,\calA 1} + A_{\calA 2} \,\#{\mathcal E}_{\,\calA 2}  \vspace{6pt}\\
\#{\mathcal H}_{\,\calA} = \displaystyle{\sqrt{\frac{\epso}{\muo}} \sqrt{\frac{\eps_\calA}{\mu_\calA}
}  \le A_{\calA 1} \,\#{\mathcal H}_{\,\calA 1} + A_{\calA 2} \,\#{\mathcal H}_{\,\calA 2} \ri  }
\end{array}
\right\},
\end{equation}
where the vectors
\begin{equation}
\left.
\begin{array}{l}  \l{A_p2}
\#{\mathcal E}_{\,\calA 1} =   \alpha_{\calA 1}  \, \ux
+ \kappa_R \, \uy +  i q  \, \uz \vspace{6pt}\\
\#{\mathcal E}_{\,\calA 2} = 
- \alpha_{\calA 2}  \, \ux
+ \kappa_L \, \uy -  i q  \, \uz
 \vspace{6pt}\\
\#{\mathcal H}_{\,\calA 1} = \displaystyle{
-i  \alpha_{\calA 1}  \, \ux
- i \kappa_R \, \uy +
q  \, \uz
   }
 \vspace{6pt}\\
\#{\mathcal H}_{\,\calA 2} = 
\displaystyle{
-i  \alpha_{\calA 2}  \, \ux +
 i \kappa_L \, \uy +
q  \, \uz
   }
\end{array}
\right\}\,.
\end{equation}


In the  half-space $z<0$, the Maxwell curl postulates yield
\begin{equation}
 \l{MP_B}
\left.
\begin{array}{l}
\#k_{\, \calB} \times \#{\mathcal E}_{\,\calB} - \omega \muo \#{\mathcal H}_{\,\calB} = \#0\vspace{4pt} \\
\#k_{\, \calB} \times \#{\mathcal H}_{\,\calB}  + \omega \epso \eps_\calB \#{\mathcal E}_{\,\calB} = \#0
\end{array}
\right\}\,,
\end{equation}
where the wave vector
 \begin{equation} \l{kB}
\#k_{\, \calB} = \ko \le q \, \ux - i \alpha_\calB \uz \, \ri,
\end{equation}
and the scalar
\begin{equation} \l{b_decay_const}
\alpha_\calB = \sqrt{q^2 -  \eps_\calB }
\end{equation}
satisfies the inequality  $\mbox{Re} \lec \alpha_\calB \ric > 0$ for surface-wave propagation.  Hence the field-phasor amplitudes are given as
\begin{equation} \l{B_p}
\left.
\begin{array}{l}
\#{\mathcal E}_{\,\calB} = A_{\calB 1} \,\uy + A_{\calB 2} \,\le i \alpha_\calB \ux + q \uz \ri \vspace{6pt}\\
\#{\mathcal H}_{\,\calB} = \displaystyle{\sqrt{\frac{\epso}{ \muo}} \les
 A_{\calB 1}\, \le i \alpha_\calB \ux + q \uz \ri - A_{\calB 2} \,  \eps_\calB\uy \, \ris  }
\end{array}
\right\}.
\end{equation}

 The scalars 
 $A_{\calA 1} $ and $A_{\calA 2} $ in Eqs.~\r{A_p}, and $A_{\calB 1} $ and $A_{\calB 2} $ in Eqs.~\r{B_p}, as well as
the relative wave number $q$,
 are determined by enforcing boundary conditions
 across the planar interface $z=0$, as follows.
 The continuity of tangential  components of the electric and magnetic field
 phasors across the planar interface  $z=0$ imposes
 the four conditions \c{Chen}
\begin{equation} \l{4cond}
 \left.
 \begin{array}{l}
  \alpha_{\calA 1}   A_{\calA 1} - \alpha_{\calA 2}  A_{\calA 2}  = i \alpha_\calB  A_{\calB 2} \vspace{6pt} \\
  \kappa_R A_{\calA 1} + \kappa_L  A_{\calA 2}  =  A_{\calB 1} \vspace{6pt} \\
\displaystyle{- 
\sqrt{\eps_\calA}
\le  \alpha_{\calA 1}   A_{\calA 1}
  + \alpha_{\calA 2} 
  A_{\calA 2} \ri = \alpha_{\calB }  \sqrt{\mu_\calA}  A_{\calB 1}}
   \vspace{6pt} \\
 \displaystyle{
 \sqrt{\eps_\calA }
\le   \kappa_R   A_{\calA 1}
  - \kappa_L 
  A_{\calA 2} \ri 
 = - i  \eps_{\calB} \sqrt{\mu_\calA}  A_{\calB 2}}
 \end{array}
 \right\}.
 \end{equation}
The four conditions \r{4cond} may be represented compactly as
\begin{equation} \l{M0}
\les M \ris \. \les \begin{array}{c}
 A_{\calA 1} \\
  A_{\calA 2} \\
   A_{\calB 1} \\
    A_{\calB 2}
\end{array}
 \ris =  \les \begin{array}{c}
 0 \\
  0 \\
   0 \\
    0
\end{array}
 \ris,
\end{equation}
wherein the 4$\times$4 matrix $\les M \ris$ must be singular for  surface-wave propagation.
The dispersion equation $\det \les M \ris = 0$ reduces to the equation
\begin{eqnarray} \l{DE}
&&
2 \sqrt{\eps_\calA \mu_\calA} \le \alpha_{\calA 1} \alpha_{\calA 2} \eps_\calB + \kappa_L \kappa_R \alpha^2_{\calB} \ri
+ \alpha_{\calB} \le \kappa_L \alpha_{\calA 1}  + \kappa_R \alpha_{\calA 2} \ri \le \eps_\calA + \eps_\calB \mu_\calA \ri = 0,
\end{eqnarray}
from which $q$ may be extracted, generally by numerical means. Once $q$ is known,  relative values of the 
four
scalars  $A_{\calA 1,2} $ and $A_{\calB 1,2} $ can be determined from Eq.~\r{M0}  by straightforward algebraic manipulations.

\subsection{Numerical studies} \label{ns1}

For our numerical studies we fix partnering material $\calA$ by selecting 
the relative constitutive parameters
$\eps_\calA = 2.6724 - 0.0007 i $, $\xi_\calA = 0.0652 + 0.0005 i$ and $\mu_A = 0.9642 + 0.0001 i $. These parameters prescribe a 
homogenized composite material that arises from blending together a realistic isotropic chiral material with an active dielectric material, namely a rhodamine mixture \c{ML_JO}. The constitutive parameters of the homogenized composite material were estimated using the Bruggeman formalism \c{MAEH}. For material $\calA$, the relative  wave numbers are $\kappa_L = 1.5401 - 0.0006 i $ and $\kappa_R =1.6705 + 0.0004 i$. Therefore, if material $\calA$ were unbounded it would simultaneously support the amplification of left circularly-polarized  plane waves and the attenuation of right circularly-polarized  plane waves \c{ML_JO}.
Parenthetically, in a similar  vein,  anisotropic dielectric materials can be engineered that  amplify plane waves of
one
 linearly-polarized state but attenuate plane waves of the other linearly-polarized state \c{ML_PRA}. 
 
\begin{figure}[!htb]
\begin{center}
\includegraphics[width=12.5cm]{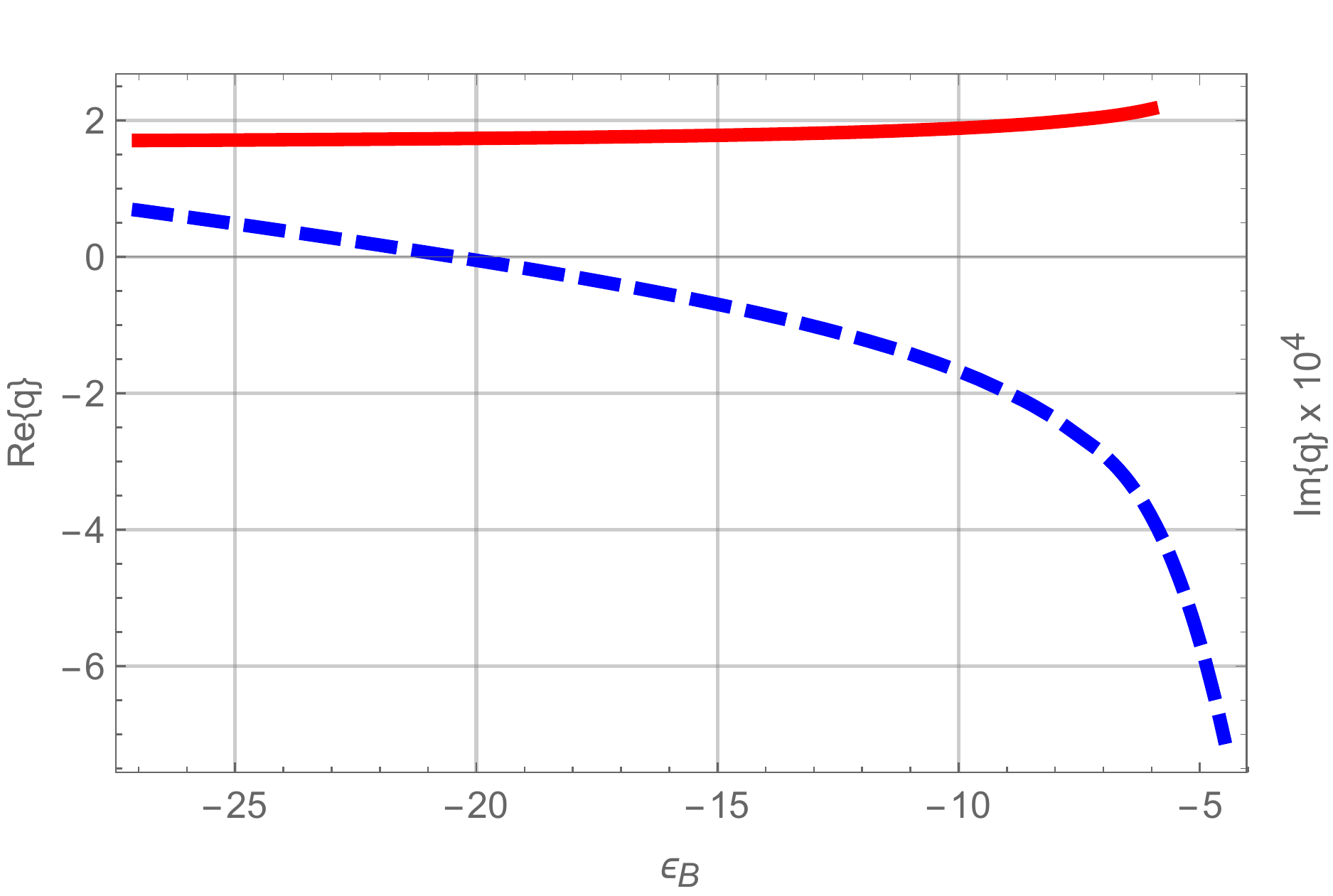}
\end{center}
 \caption{${\rm Re}\lec q \ric$ (red solid  curve)  and ${\rm Im}\lec q \ric$ (blue dashed  curve)    plotted against  $\eps_\calB $.
 } \label{Fig1}
\end{figure}

Let us begin by considering the case wherein partnering material $\calB$ is characterized by a real-valued relative permittivity with $\eps_\calB < 0$, That is, material $\calB$ behaves like an idealized plasmonic material
and the corresponding surface waves are akin to Fano waves \c{ESW_book,Fano}. The dispersion equation 
\r{DE}
then yields one (complex-valued) $q$ solution. The real and imaginary parts of the relative wave number $q$ are plotted 
against relative permittivity $\eps_\calB$ in Fig.~\ref{Fig1}. While the real part of $q$ remains almost constant as $\eps_\calB$ is increased, the imaginary part of $q$ undergoes a much more dramatic change. In particular, $\mbox{Im} \lec q \ric$ is positive valued for $\eps_\calB < -21$ but is negative valued for $\eps_\calB > -21$. Therefore, for sufficiently small values of
$\eps_\calB$ the surface wave is attenuated as it propagates whereas for larger values of 
$\eps_\calB$ the surface wave is amplified as it propagates.

\begin{figure}[!htb]
\begin{center}
\includegraphics[width=12.5cm]{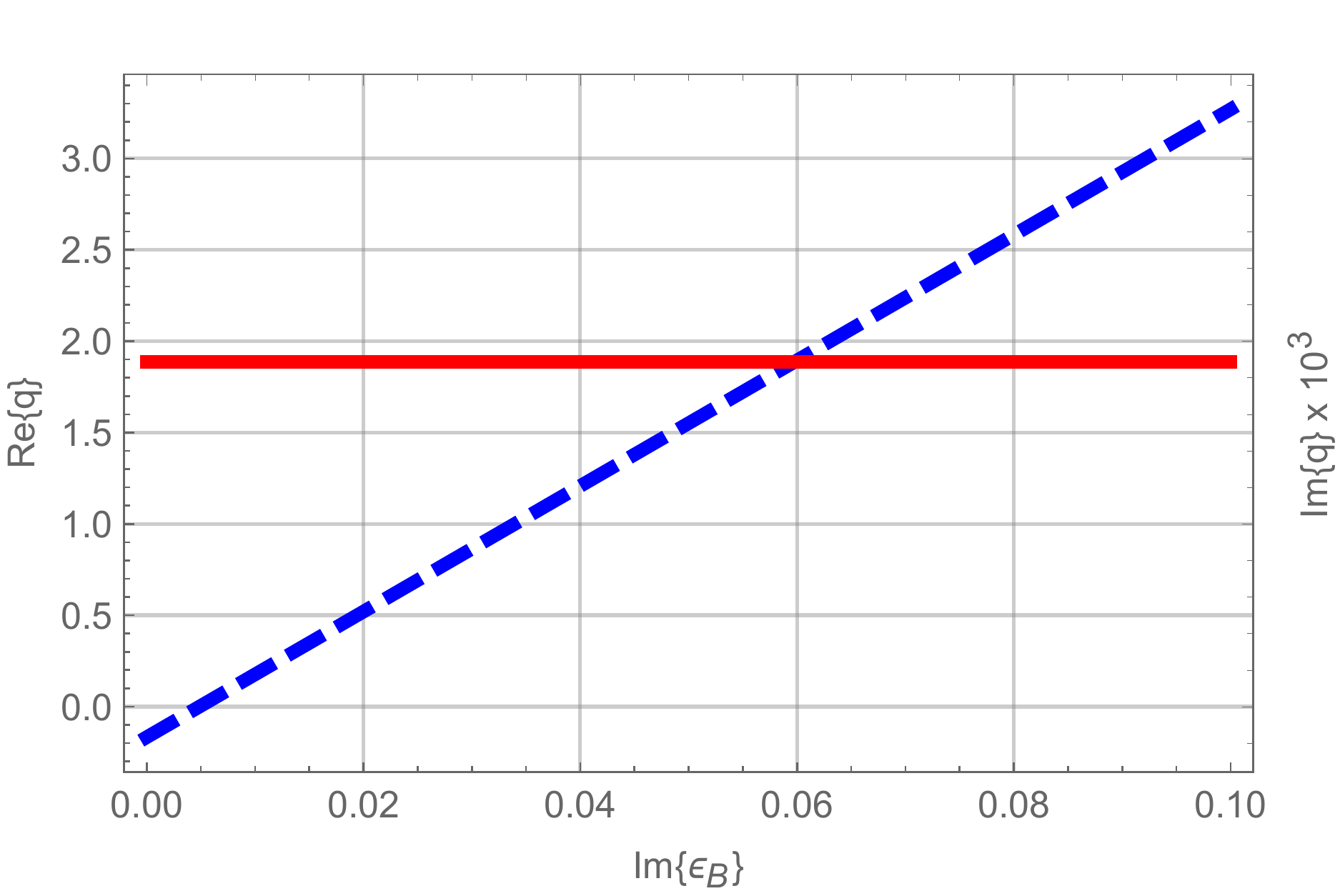}
\end{center}
 \caption{ ${\rm Re}\lec q \ric$ (red solid  curve)  and ${\rm Im}\lec q \ric$ (blue dashed  curve)    plotted against
   $\mbox{Im} \lec \eps_\calB \ric$ with  $\mbox{Re} \lec \eps_\calB \ric = - 10$.
 } \label{Fig2}
\end{figure}

We explore this behaviour further, in a more realistic setting, by supposing that partnering material $\calB$
 exhibits a small degree of dissipation. That is, we fix $\mbox{Re} \lec \eps_\calB \ric = - 10$ and consider $\mbox{Im} \lec \eps_\calB \ric > 0 $.
Accordingly,  material $\calB$ behaves like a realistic plasmonic material and the corresponding surface waves are akin to SPP waves \c{BGprA,BGprB}.
As for the case represented in Fig.~\ref{Fig1}, the dispersion equation 
\r{DE}
delivers a single (complex-valued) $q$ solution. The real and imaginary parts of the relative wave number $q$ are plotted 
against the imaginary part of the relative permittivity $\eps_\calB$ in Fig.~\ref{Fig2}. 
The real part of $q$ is almost independent of $\mbox{Im} \lec \eps_\calB \ric$ over the range considered in Fig.~\ref{Fig2}. 
In contrast, for $\mbox{Im} \lec \eps_\calB \ric < 0.005$ we have  $\mbox{Im} \lec q \ric < 0$
whereas for $\mbox{Im} \lec \eps_\calB \ric > 0.005$ we have  $\mbox{Im} \lec q \ric > 0$. Therefore, 
when the degree of dissipation exhibited by material $\calB$ is sufficiently small the surface wave is amplified, but 
when the degree of dissipation exhibited by material $\calB$ is larger the surface wave is attenuated.

\begin{figure}[!htb]
\begin{center}
\includegraphics[width=12.5cm]{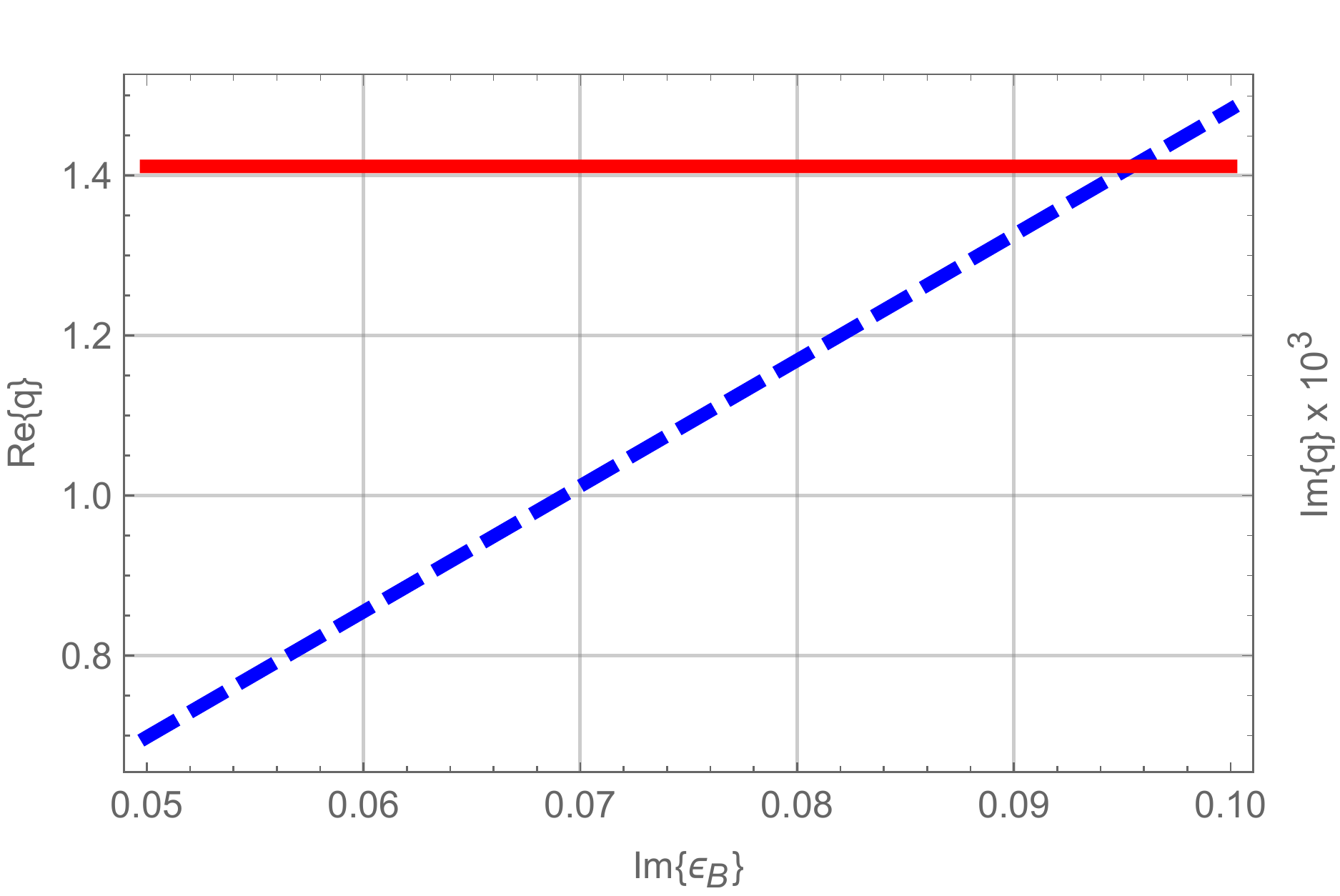}
\end{center}
\caption{As Fig.~\ref{Fig2} except that  $\mbox{Re} \lec \eps_\calB \ric =  10$.
 } \label{Fig3}
\end{figure}

Next we turn to the case where $\mbox{Re} \lec \eps_\calB \ric > 0$. 
Specifically, let $\mbox{Re} \lec \eps_\calB \ric = 10$.
Notice that in this case
partnering material $\calB$ does not behave like a plasmonic material and the corresponding surface waves
 are not at  all akin to 
SPP waves. In Fig.~\ref{Fig3}, the real and imaginary parts of $q$ are plotted against 
$\mbox{Im} \lec \eps_\calB \ric$, for the single (complex-valued) solution emerging from the dispersion equation 
\r{DE} for $\mbox{Im} \lec \eps_\calB \ric > 0.05$. The real part of $q$ is almost independent of $\mbox{Im} \lec \eps_\calB \ric$ whereas the imaginary part of $q$ increases approximately  linearly as $\mbox{Im} \lec \eps_\calB \ric$ increases.
Since $\mbox{Im} \lec q \ric > 0 $, the corresponding surface wave is attenuated over the range of values of $\mbox{Im} \lec \eps_\calB \ric$ considered. If $ 0 \leq \mbox{Im} \lec \eps_\calB \ric < 0.05$ then no solutions emerge from the dispersion equation 
\r{DE}. In particular, no solutions are found when material $\calB$ is nondissipative and  $\eps_\calB > 0$, a result that is consistent with an
earlier study \c{Fantino}.

\section{Isotropic chiral  material/anisotropic dielectric  material interface} \l{aniso_sec}

Now we extend the canonical boundary-value problem  presented in Sec.~\ref{iso_sec} by replacing 
the
isotropic dielectric material, i.e., partnering material $\calB$, with an anisotropic dielectric material. Partnering material $\calA$ is still taken to be an isotropic chiral material, per the Tellegen constitutive relations \r{cr}.
To be specific, partnering material $\calB$ is taken to be a uniaxial dielectric material characterized by the relative permittivity dyadic \c{Chen}
\begin{equation}
\=\eps_\mathcal{B} = \eps_\mathcal{B}^{\rm s} \=I + \le
\eps_\mathcal{B}^{\rm t} - \eps_\mathcal{B}^{\rm s} \ri \,
\hat{\#{u}}_{\rm } \, \hat{\#{u}}_{\rm }\,, \l{Ch4_eps_uniaxial}
\end{equation}
where $\=I=\ux\ux+\uy\uy+\uz\uz$ is the identity dyadic.
The
optic axis  of material $\calB$ lies in the $xy$ plane, 
oriented at angle $\psi$
with respect to the direction of surface-wave propagation; i.e,
\begin{equation}
\hat{\#{u}}_{\rm } = \cos \psi \, \ux +  \sin \psi \, \uy.
\end{equation}
The dyadic $\=\eps_\mathcal{B}$ has two eigenvalues: $\eps_\mathcal{B}^{\rm s}$  which governs the propagation of \textit{ordinary}
plane waves and $\eps_\mathcal{B}^{\rm t}$  which, together with $\eps_\mathcal{B}^{\rm s}$ ,
 governs the propagation of \textit{extraordinary}
plane waves  \cite{BW}.

\subsection{Theory}

In the  half-space $z>0$,  Eqs.~\r{MP_A}--\r{A_p2} continue to hold.
In the  half-space $z<0$, the Maxwell curl postulates yield
\begin{equation} \l{xMP_A}
\left.
\begin{array}{l}
\#k_{\, \calB} \times \#{\mathcal E}_{\,\calB} = \omega \muo \#{\mathcal H}_{\,\calB} \vspace{4pt} \\
\#k_{\, \calB} \times \#{\mathcal H}_{\,\calB} = - \omega  \epso \=\eps_{\,\calB} \. \#{\mathcal E}_{\,\calB}
\end{array}
\right\}\,,
\end{equation}
where the wave vector $\#k_{\, \calB}$ has the form given in Eq.~\r{kB}.
The combination of  Eqs.~\r{xMP_A} and Eq.~\r{kB} yields a biquadratic dispersion relation
 for $\alpha_\calB$.
For surface-wave progation the two $\alpha_\calB$ roots with non-negative real parts are prescribed; these are
\begin{equation} \l{xa_decay_const}
\left.
\begin{array}{l}
\alpha_{\calB 1} = \sqrt{q^2 -\eps_\calB^s} \vspace{8pt}\\
\alpha_{\calB 2} = \displaystyle{
\sqrt{ \eps^t_\calB \les
q^2 \le \frac{ \cos^2 \psi}{\eps^s_\calB} +
\frac{\sin^2 \psi}{\eps^t_\calB} \ri -1 \ris}}
\end{array}
\right\}\,.
\end{equation}
Hence the field phasor amplitudes for $z<0$  are
\begin{equation}
\left.
\begin{array}{l} \l{B_p2}
\#{\mathcal E}_{\,\calB} = A_{\calB 1}\, \#{\mathcal E}_{\,\calB 1} + A_{\calB 2} \,\#{\mathcal E}_{\,\calB 2}  \vspace{6pt}\\
\#{\mathcal H}_{\,\calB} = \displaystyle{\sqrt{\frac{\epso}{\muo}}  \le A_{\calB 1} \,\#{\mathcal H}_{\,\calB 1} + A_{\calB 2} \,\#{\mathcal H}_{\,\calB 2} \ri  }
\end{array}
\right\},
\end{equation}
where the vectors
\begin{equation}
\left.
\begin{array}{l}
\#{\mathcal E}_{\,\calB 1} =   i \alpha_{\calB 1} \sin \psi \, \ux
- i \alpha_{\calB 1} \cos \psi \, \uy + q \sin \psi \, \uz \vspace{6pt}\\
\#{\mathcal E}_{\,\calB 2} =  \alpha^2_{\calB 1} \cos \psi \, \ux
-  \eps^s_\calB \sin \psi \, \uy - i q \alpha_{\calB 2} \cos \psi \, \uz \vspace{6pt}\\
\#{\mathcal H}_{\,\calB 1} = \displaystyle{
 \alpha^2_{\calB 1} \cos \psi \, \ux
- \eps^s_\calB \sin \psi \, \uy -
i  \alpha_{\calB 1} q \cos \psi \, \uz
   }
 \vspace{6pt}\\
\#{\mathcal H}_{\,\calB 2} = \displaystyle{ -  i \alpha_{\calB 2} \eps^s_\calB \sin \psi \, \ux
+ i \alpha_{\calB 2} \eps^s_\calB \cos \psi \, \uy -
 \eps^s_\calB q \sin \psi \, \uz
   }
\end{array}
\right\}\,.
\end{equation}

As in   Sec.~\ref{iso_sec},  the scalars 
 $A_{\calA 1} $ and $A_{\calA 2} $ in Eqs.~\r{A_p} , and $A_{\calB 1} $ and $A_{\calB 2} $ in Eqs.~\r{B_p2}, as well as
the wave number $q$,
 are determined by enforcing boundary conditions
 across the planar interface $z=0$.  To this end,
 the continuity of tangential  components of the electric and magnetic field
 phasors across the planar interface  $z=0$ yields
 the four conditions \c{Chen}
\begin{equation} \l{4cond2}
 \left.
 \begin{array}{l}
   \alpha_{\calA 1}   A_{\calA 1} - \alpha_{\calA 2}  A_{\calA 2}  =  i \alpha_{\calB 1} \sin \psi A_{\calB 1} 
  + \alpha^2_{\calB 1} \cos \psi A_{\calB 2}
  \vspace{6pt} \\
  \kappa_R A_{\calA 1} + \kappa_L  A_{\calA 2}  =  
 -   i \alpha_{\calB 1} \cos \psi A_{\calB 1} 
  - \eps^{\rm s}_{\calB } \sin \psi A_{\calB 2}
   \vspace{6pt} \\
\displaystyle{- i
\sqrt{\frac{\eps_\calA}{\mu_\calA}}
\le  \alpha_{\calA 1}   A_{\calA 1}
  + \alpha_{\calA 2} 
  A_{\calA 2} \ri = \sqrt{\frac{\epso}{\muo}} \le \alpha^2_{\calB 1} \cos \psi A_{\calB 1} 
  - i  \alpha_{\calB 2} \eps^{\rm s}_{\calB } \sin \psi A_{\calB 2} \ri
 }
   \vspace{6pt} \\
 \displaystyle{-i
 \sqrt{\frac{\eps_\calA}{\mu_\calA}}
\le   \kappa_R   A_{\calA 1}
  - \kappa_L 
  A_{\calA 2} \ri 
 =  \sqrt{\frac{\epso}{\muo}} \le -  \eps^{\rm s}_{\calB }  \sin \psi A_{\calB 1} 
  + i  \alpha_{\calB 2} \eps^{\rm s}_{\calB } \cos \psi A_{\calB 2} \ri }
 \end{array}
 \right\},
 \end{equation}
which are  represented compactly as
\begin{equation} \l{N0}
\les N \ris \. \les \begin{array}{c}
 A_{\calA 1} \\
  A_{\calA 2} \\
   A_{\calB 1} \\
    A_{\calB 2}
\end{array}
 \ris =  \les \begin{array}{c}
 0 \\
  0 \\
   0 \\
    0
\end{array}
 \ris.
\end{equation}
The 4$\times$4 matrix $\les N \ris$ must be singular for  surface-wave propagation.
The dispersion equation $\det \les N \ris = 0$ reduces to the equation
\begin{eqnarray} \l{DE2}
&&
\alpha_{\calB 1} \les  2 \sqrt{\eps_\calA \mu_\calA} \le \kappa_L \kappa_R \alpha^3_{\calB 1} + \eps^s_\calB \alpha_{\calA 1} \alpha_{\calA 2}
\alpha_{\calB 2} \ri + \alpha_{\calB 1} \le \kappa_L \alpha_{\calA 1} + \kappa_R \alpha_{\calA 2} \ri \le \eps_\calA \alpha_{\calB 1} + \eps^s_\calB
\mu_\calA \alpha_{\calB 2} \ri  \ris \cos^2 \psi \nonumber \\ &&  -\eps^s_\calB  \left\{  \les  2 \sqrt{\eps_\calA \mu_\calA}  \le \kappa_L \kappa_R \alpha_{\calB 1} \alpha_{\calB 2} + \eps^s_\calB \alpha_{\calA 1}
\alpha_{\calA 2} \ri + \le \kappa_L \alpha_{\calA 1} + \kappa_R \alpha_{\calA 2} \ri \le \eps_\calA \alpha_{\calB 1} + \eps^s_\calB \mu_\calA \alpha_{\calB 2}
\ri \ris  \sin^2 \psi \right. \nonumber \\ && \left.
-  \sqrt{\eps_\calA \mu_\calA}  \le \kappa_L \alpha_{\calA 1} - \kappa_R \alpha_{\calA 2} \ri \alpha_{\calB 1} \le \alpha_{\calB 1} - \alpha_{\calB 2} \ri \sin 2 \psi 
\right\} = 0.
\end{eqnarray}
After extracting  $q$ from Eq.~\r{DE2}, generally by numerical means, the  relative values of the 
four
scalars  $A_{\calA 1,2} $ and $A_{\calB 1,2} $ can be determined from Eq.~\r{N0} by straightforward algebraic manipulations.

 \subsection{Numerical studies}\label{ns2}

\begin{figure}[!htb]
\begin{center}
\includegraphics[width=12.5cm]{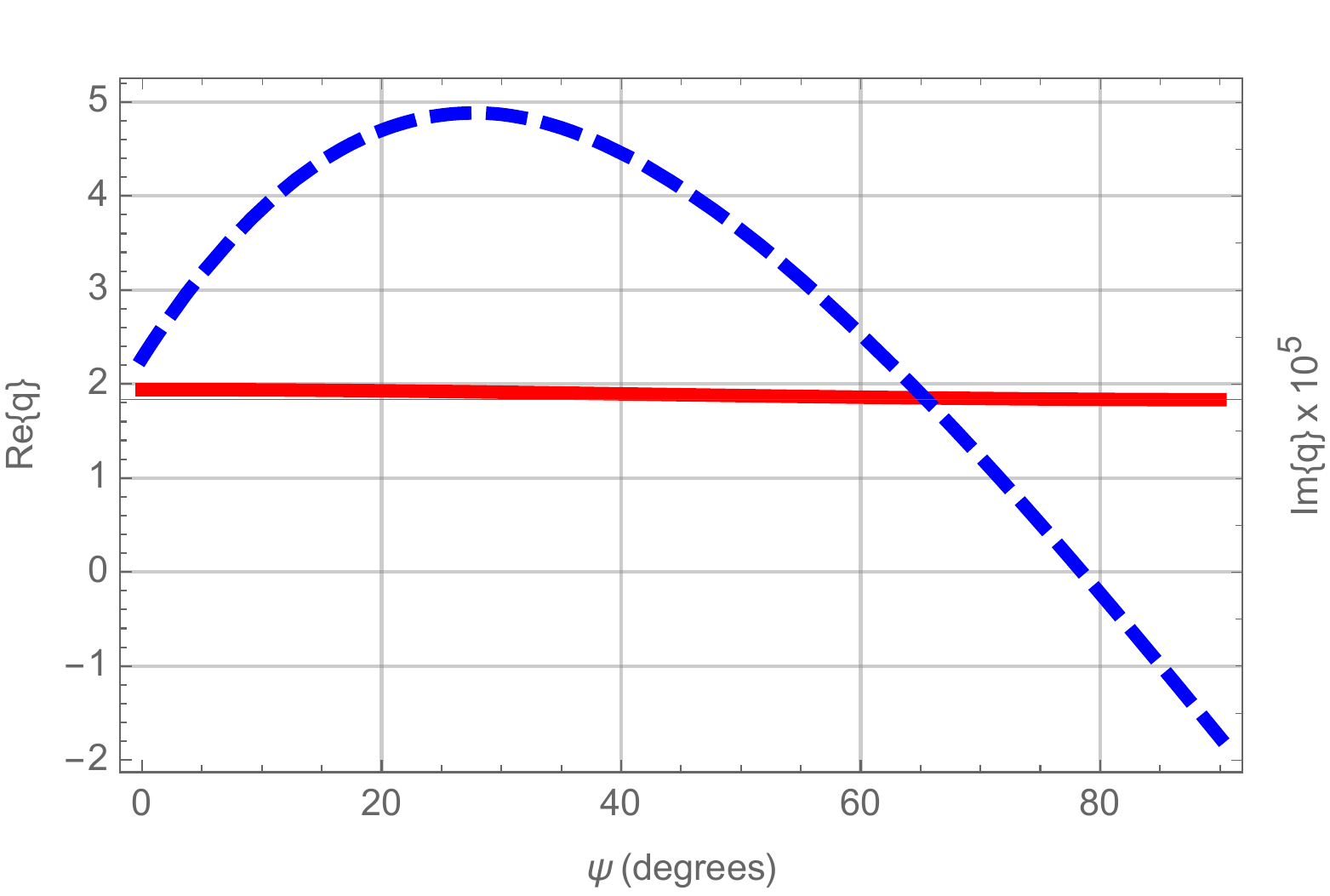}
\end{center}
 \caption{${\rm Re}\lec q \ric$ (red solid  curve)  and ${\rm Im}\lec q \ric$ (blue dashed  curve)    plotted against orientation
 angle   $\psi$ with $ \epsilon^s_\calB = -12 + 0.0045 i$ and $\epsilon^t_\calB = -8 + 0.005i$.
 } \label{Fig4}
\end{figure}

As in Sec.~\ref{ns1},   we fix partnering material $\calA$ by selecting 
the relative constitutive parameters
$\eps_\calA = 2.6724 - 0.0007 i $, $\xi_\calA = 0.0652 + 0.0005 i$ and $\mu_A = 0.9642 + 0.0001 i $. 
Let us start with the case where
the relative permittivity parameters of material $\calB$, namely $\eps^s_\calB$ and $\eps^t_\calB$,  
have negative-valued real parts and positive-valued imaginary parts. 
Specifically, let $\eps^s_\calB = -12 + 0.0045 i $ and $\eps^t_\calB= - 8 + 0.005i$.
Thus, material $\calB$ behaves like an anisotropic plasmonic material that exhibits a modest degree of dissipation. The corresponding surface waves are akin to SPP waves.
 In Fig.~\ref{Fig4}, the real and imaginary parts of $q$ are plotted against orientation angle $\psi$
 for the solitary (complex-valued) $q$ solution emerging from the dispersion equation 
\r{DE2}. The real part of $q$ is almost independent of  $\psi$. In contrast, the imaginary part of $q$  oscillates as $\psi$ increases such that $\mbox{Im} \lec q \ric > 0$ for $\psi \in \le 0^\circ, 64^\circ \ri \cup
\le 170^\circ, 180^\circ \ri$ and $\mbox{Im} \lec q \ric < 0$ 
for $\psi \in \les 64^\circ, 170^\circ  \ris$. Therefore, for certain orientations of material $\calB$ the surface wave is attenuated as it propagates while for other orientations 
it is amplified as it propagates.
Furthermore, at two specific orientations of material $\calB$ the surface wave propagates with neither attenuation nor amplification.
 A similar  simultaneous  amplification/attenuation phenomenon has been reported for  SPP waves guided by the planar interface of a metal and a uniaxial dielectric material \c{ML_SPP_JO}.
 
\begin{figure}[!htb]
\begin{center}
\includegraphics[width=12.5cm]{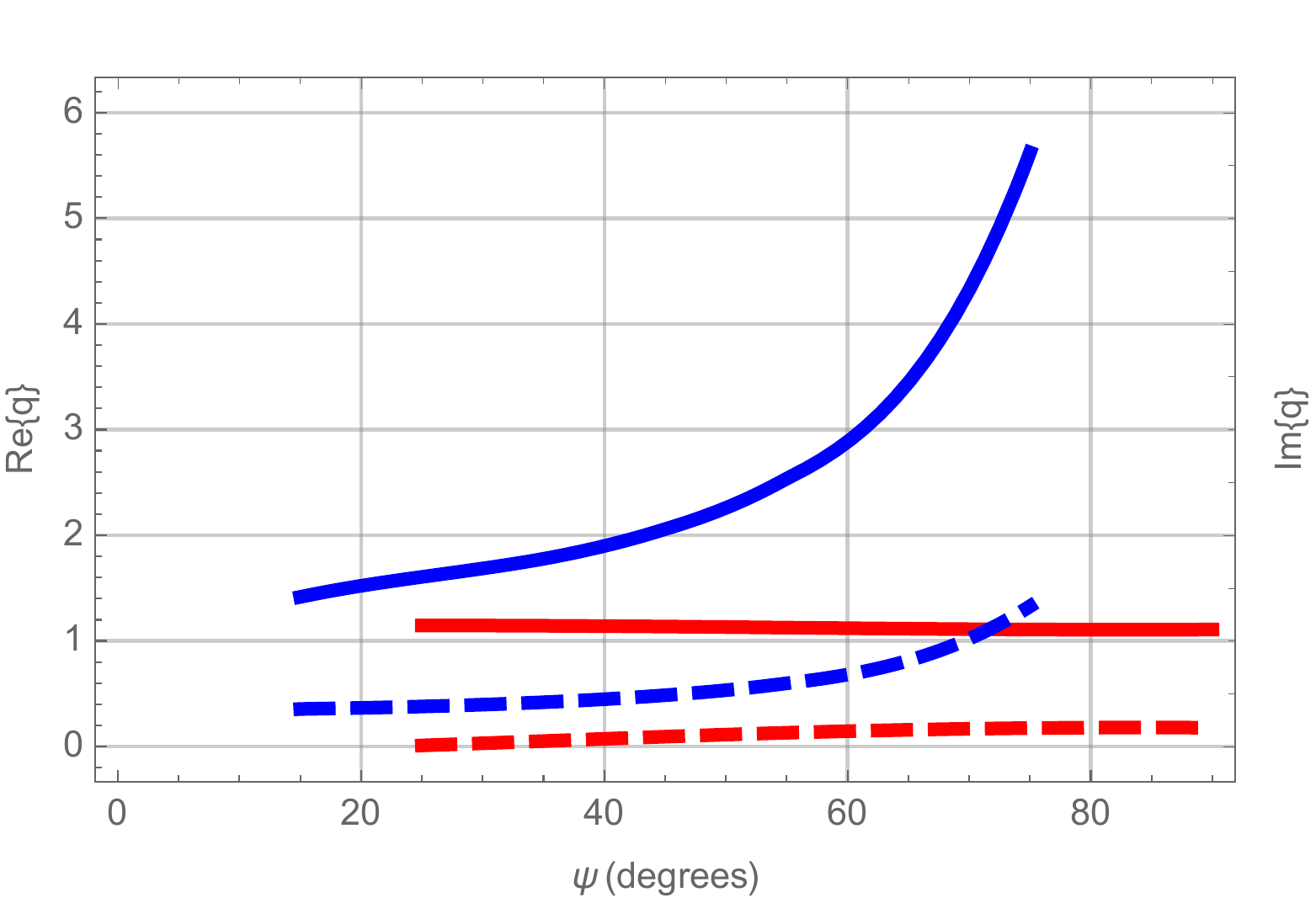}
\end{center}
\caption{As Fig.~\ref{Fig4} except that $ \epsilon^s_\calB = 2 + i$ and $\epsilon^t_\calB = 3 + 1.5i$.
 } \label{Fig5}
\end{figure}

 Lastly, we turn to the case  where
the relative permittivity parameters of material $\calB$
have positive-valued real parts and positive-valued imaginary parts. 
Specifically, let $\eps^s_\calB = 2 +  i $ and $\eps^t_\calB= 3 + 1.5i$.
Thus, material $\calB$ behaves like an anisotropic dielectric material that exhibits dissipation. The corresponding surface waves are akin to Dyakonov waves \c{Dyakonov88,Takayama_exp}.
Unlike the case represented in Fig.~\ref{Fig4}, the dispersion equation \r{DE2} here yields two, one or no $q$ solutions, depending upon the value of $\psi$. 
The real and imaginary parts of $q$ are  plotted against $\psi$ for all solutions in Fig.~\ref{Fig5}.
  Specifically, we found two $q$ solutions for $23^\circ < \psi < 76^\circ$, one $q$ solution
for $14^\circ < \psi < 23^\circ$ and $76^\circ < \psi < 90^\circ$, and no $q$ solution at all for $0^\circ < \psi < 14^\circ$. The existence of two solutions signifies a marked difference from the case for conventional Dyakonov waves \c{ESW_book}. 
In the case represented in  Fig.~\ref{Fig5} all surface-wave solutions are attenuated regardless of the orientation angle $\psi$, which contrasts with 
a   simultaneous  amplification/attenuation phenomenon that has been reported for  Dyakonov waves guided by the planar interface of an isotropic dielectric material and a uniaxial dielectric material \c{ML_Dyakonov}. Also, the existence of two surface waves at each orientation   for $23^\circ < \psi < 76^\circ$  in  Fig.~\ref{Fig5} contrasts with the usual case for Dyakonov waves in which the dispersion equation admits only one solution for each propagation direction \c{DSWreview}.

\section{Closing remarks}\label{dcr}

Electromagnetic surface waves supported by an isotropic chiral material were investigated numerically via the associated canonical boundary-value problem. In the case where
 the partnering material was an isotropic dielectric material, surface waves akin to SPP waves were excited;
 while in the case where the partnering material was a  uniaxial dielectric material,
surface waves akin to Dyakonov waves and SPP waves were excited.  By selecting an isotropic chiral material capable of  simultaneously supporting attenuation and amplification of plane waves, depending upon circular polarization state, the amplitudes of the SPP-like waves 
 could be either amplified or attenuated, depending upon the relative permittivity of the isotropic dielectric partnering material
 or the direction of propagation relative to the optic axis of the uniaxial dielectric partnering material.
Therefore, wider opportunities for   surface-wave propagation  are supported by isotropic chiral materials, as compared to achiral materials in analogous scenarios.

\vspace{10mm}

\noindent {\bf Acknowledgment} JN is supported by  a  Deans' Vacation Scholarship (University of Edinburgh).

\newpage

\newpage

\newpage

\newpage

\newpage

\end{document}